\documentclass[useAMS,usenatbib,usegraphicx]{mn2e}

\bibpunct{(}{)}{;}{a}{}{,}

\newcommand{\ignore}[1]{}

\newcommand{\rv}[1]{#1}

\usepackage{mathptmx}
\usepackage{amssymb}
\usepackage{color}
\title[Stability of Orbits in the Broad Line Region]
{Stability of Cloud Orbits in the Broad Line Region of Active
  Galactic Nuclei}
\author[M. Krause et al.]{Martin Krause$^{1,2}$\thanks{E-mail:
krause@mpe.mpg.de,mkrause@usm.lmu.de}, 
Andreas Burkert$^{1,2,3}$, Marc Schartmann$^{1,2}$\\
$^1$Max-Planck-Institut f\"ur Extraterrestrische Physik, Giessenbachstrasse, 85748 Garching, Germany \\
$^2$Universit\"atssternwarte M\"unchen, Scheinerstr.~1, 81679 M\"unchen, Germany\\
$^3$Max-Planck-Fellow\\}
\begin{document}

\date{Accepted \date. Received 2009 June 25?}

\pagerange{\pageref{firstpage}--\pageref{lastpage}} \pubyear{2008}

\maketitle

\label{firstpage}

\begin{abstract}
We investigate the global dynamic stability of spherical clouds in the Broad Line
Region (BLR) of Active Galactic Nuclei (AGN), exposed to radial
radiation pressure, gravity of the central black hole (BH), and
centrifugal forces assuming the clouds adapt their size according to the local pressure.
We consider both, isotropic and anisotropic light
sources. In both cases, stable orbits exist also for very
sub-Keplerian rotation 
for which the
radiation pressure contributes substantially to the force budget.
We demonstrate that highly eccentric, very sub-Keplerian stable orbits 
may be \rv{found.
This} gives further support for the model of 
\citet{Marcea08}, \rv{who pointed out that black hole masses might be
significantly underestimated if radiation pressure is neglected. That
model improved} the agreement between
black hole masses derived in certain active galaxies based on BLR dynamics,
and black hole masses derived by other means in other galaxies by
inclusion of a luminosity dependent term. For anisotropic
illumination, \rv{energy is conserved for averages over long time
intervals, only, but not for individual orbits. This leads to Rosetta
orbits that are systematically less extended in the direction of
maximum radiation force. Initially isotropic relatively low column
density systems would therefore turn into a disk when an anisotropic
AGN is switched on.}
\end{abstract}

\section{Introduction}\label{intro}
Recently, the BLR has received much attention, because the line
widths, in combination with the size of the region, measured for example by
reverberation mapping, allows a determination of the mass of the
central super-massive black hole
\citep[e.g.][]{Benea09,Net09}, which is essential if one
wants to place AGN in the context of general galaxy evolution.  From
such studies, one finds that AGN generally have normal sized black
holes\rv{, as expected for the size of their host galaxy,} and
that most galaxies must have been active for one or more times in
their life. The measurement of the black hole mass is affected by the
relative importance of radiation pressure and gravity, which should
dominate the force budget. This is
connected to the properties of individual BLR clouds and the dynamical
configuration of the system \citep{Marcea08}.

The BLR is a standard AGN ingredient \citep[e.g.][]{Ost88,Pet97,Net08}: It is
located inside the obscuring torus, above and below the accretion
disk, and at a distance of fractions of a parsec from the
super-massive black hole.
The line emission is powered by
photoionisation by a broad-band continuum due to the innermost parts 
of the accretion disk. Photoionisation also governs the
thermodynamics of these clouds: They have a stable equilibrium
temperature of order $10^4$~K, independent of their locations. 
The critical densities of
observed and suppressed emission lines and pressure equilibrium
considerations \citep{FE84} constrain the number densities
in the clouds to $n_\mathrm{cl}=10^{10\pm1}$~cm$^{-3}$. 
Signatures from partially
ionised zones are typically observed. Therefore, the Str\"omgren
depth is a lower limit of the cloud size. An upper limit may be found
by requiring the lines to be smooth, which demands a certain minimum
on the number of individual entities. This results in cloud sizes of
about $r_\mathrm{cl} = 10^{12\pm1}$~cm \rv{\citep[][and references therein]{Laorea06}}. 
In agreement with this, detailed photoionisation models \citep{KK81} constrain the column
densities to $N_\mathrm{cl}> 10^{22}$~cm$^{-2}$.
Because BLR clouds are optically thick, high ionisation lines are
produced only by the illuminated surface of the clouds. In such lines,
one observes generally the far side of the BLR, wherefore outflows
manifest themselves as redshifts, inflows as blueshifts.

Much information has also been gathered on the dynamical
state of the BLR. Inclination matters: the statistics of line widths
excludes Keplerian rotation in a flat disk, but is 
instead consistent with a thick disk configuration with $v/\sigma \approx 2.5$, where
$v$ is the rotational velocity and $\sigma$ the turbulent one \citep{Ost78}. 
For radio-loud AGN, the ratio of core to lobe power correlates with
the width of the broad lines in the sense that the systems observed at a
line-of-sight close to the jet axis have the narrower lines
\citep{WB86}.
\rv{This finding was confirmed by \citet{JMcL06}. They find that the average line width of flat
spectrum radio sources, which are seen more face on, have
narrower emission lines by about 30\% than steep spectrum radio sources.}
Spectropolarimetry of type~2 objects, where the BLR is hidden by a
dusty torus, has revealed hidden BLRs in polarised light, which
established the orientation unification scenario \citep{Ant93}.
In these objects, the BLR emission is scattered by a polar region
above the BLR. The polarisation angle is perpendicular to the system
axis, defined by the radio \rv{\citep{Smea04}} or UV/blue \citep{Borea08}
emission. The same authors find a preferentially parallel orientation
for type~1 objects, which provide an unobscured view towards the
BLR. This finding is explained by an equatorial scattering region,
much closer to the BLR. Where observed (type~1 objects), the latter
dominates the polarised emission. The equatorial scatterer is in fact
very close to the BLR and provides unambiguous evidence for a disk-like
configuration of the BLR \citep{Smea05}: Approaching and receding
parts of the BLR-disk are seen at slightly different angles by the
equatorial scatterer. This produces a noticeable difference in
polarisation angle for the blue and red parts of the line. This, and
other characteristics, has been observed by \rv{\citet{Smea04,Smea05}} for a
sample of Seyfert galaxies. They find a continuum of polarisation
properties: Zero polarisation for near pole on objects, where the
different contributions by equatorial scatterers in various directions
cancel each other; wavelength dependent parallel polarisation for
intermediate inclination type~1 objects, where parts of the BLR and the
equatorial scattering region are attenuated by the obscuring torus;
and perpendicular polarisation for type~2 objects.
\rv{Other evidence pointing to a disk-like nature of the BLR comes
  from} the existence of objects with double
peaked broad lines indicative of Keplerian motion \citep[][12\% in
their sample]{EH03}, while the majority of single peaked objects allows
for a hidden disk component, if the inclination is small, and the
contribution of the flattened part is not too strong \citep{Bonea09}.
These findings suggest that the dominant contribution to the BLR
kinematics is Keplerian rotation, followed by a turbulent component.
There is also evidence for radial motion:
There are examples in the literature for bulk outflows, as measured by
spectropolarimetry \citep[][and references therein]{Yea07}. Velocity
resolved reverberation mapping confirms the dominant Keplerian
motions, but additionally finds evidence for bulk inflow and outflow in individual
objects \citep[e.g.][]{DK96,UH96,Kol03,Benea08,Benea09,Denea09}.
\citet{Gas08} interprets the combined evidence of velocity resolved
reverberation mapping as strong evidence that the radial part of the
kinematics is dominated by inflow.

Radiation pressure due to the central parts of the accretion disk is
generally assumed to be significant in the BLR
\citep[e.g.][]{BM75,Mat86,Marcea08,Net09}. Since the dependence on distance
to the light emitter is an inverse square law, like for gravity, the
effect of radiation pressure support is to reduce the apparently
measured mass of the black hole, based on BLR kinematics. Recently, it
has been under debate, how strong this effect was
\citep{Marcea08,Marcea09,Net09,Net10}, especially
in the case of NLS1 galaxies.
It is interesting to ask in this context, which 
cloud orbits would be stable against small perturbations, and also, which
ones are compatible with the spectropolarimetric results.
In the following we perform such an analysis.

We present the cloud orbit analysis in sect.~\ref{stab}, discuss our
finding in sect.~\ref{disc} and summarise our conclusions in sect.~\ref{conc}.

\section{Dynamical stability analysis}\label{stab}
We carry out the analysis for both, an isotropic light source, and a
light source with a $\cos \theta$ dependence for the luminosity, where
$\theta$ is the polar angle. The latter case should be more realistic,
as it is the expected angular distribution for the luminosity of a thin
accretion disk. 

\subsection{Isotropic light source}
\rv{For optically thick, spherical clouds of constant mass $m$ and
central column density $N$, without internal structure, 
the force equation in spherical polar coordinates reads}:
\begin{equation}\label{eq:focl}
F= \frac{GM_\mathrm{BH} m}{R^2} 
\left( \frac{3 l}{2\sigma_\mathrm{T} N} + V^2 -1\right) \, ,
\end{equation}
where $\sigma_\mathrm{T}$ is the Thomson cross section, $l$ the
luminosity in Eddington units, $V$ the rotational velocity in Kepler
units and $M_\mathrm{BH}$ the mass of the black hole. 

Dynamical equilibrium, corresponding to circular orbits, is reached for a column density of:
\begin{equation}\label{Neq}
N =\frac{3l}{2 \sigma_\mathrm{T} (1-V^2)} \, .
\end{equation}
\begin{figure}
\centering
\includegraphics[width=.47\textwidth]{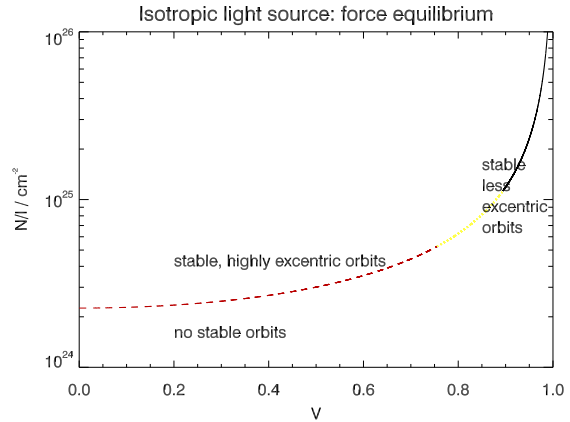}
\caption{\small  Dynamical equilibrium column density over luminosity
  in Eddington units against rotation velocity in Kepler units for an
  isotropic light source. The red, dashed part of the line corresponds
  to a maximum of the effective potential for reasonable choices of
  the parameter $s$ ($s>1$) that
  characterises the pressure profile. Stable orbits are still found in
  this case, but are highly eccentric, and are found above the red
  line. 
The yellow, dotted part
  corresponds to a stable minimum for certain values
  of $s$. The solid, black region is always a minimum of the effective
  potential, provided $s<3$, and therefore allows for orbits with low
  eccentricity, which scatter around the line.  See text for more details.}
\label{fig:dec_N}
\end{figure}
For a large range of \rv{rotational velocities $V$}, this is of order $10^{24} \,
l$~cm$^{-2}$ (compare Fig.~\ref{fig:dec_N}). 

To assess dynamical stability, we now consider perturbations to
circular orbits. Following \citet{Net08}, we assume a confining inter-cloud
medium with a pressure profile of $p(R)\propto R^{-s}$. The sound
crossing time through a BLR cloud is of order weeks, whereas the
orbiting period is many years. We therefore assume the clouds to
adjust their radius instantaneously to any change in external
pressure. A change in the clouds' cross section alters the amount
of radiation received and hence the radiation pressure support. Clouds
will be dynamically stable, if an increase (decrease) in orbital radius $R$ leads to a
net inward (outward) force, i.e. a position above (below) the
equilibrium line in Fig.~\ref{fig:dec_N}. We proceed to calculate the
perturbed locations in the $N/l - V$-diagram.

We first consider a spherically symmetric cloud initially in a circular
orbit at distance~$R$ to the black hole. Since the temperature is kept
constant by photoionisation, a change in \rv{distance} by $\Delta R>0$ will
result in a change of the particle density by $n(R+\Delta R) = n(R)
[(R+\Delta R)/R]^{-s}$. This results in a change of the column density:
\begin{equation}\label{dN}
N(R+\Delta R)=N(R) (1+\Delta R/R)^{-2s/3}\, .
\end{equation}
The corresponding change in orbital velocity for angular momentum
conserving perturbations is: 
\begin{equation}\label{dV}
V(R+\Delta R)=V(R) (1+\Delta R/R)^{-1/2}\, .
\end{equation}
The perturbed cloud will receive a restoring force, if:
\begin{equation}
N(R+\Delta R) > \frac{3l}{2\sigma_\mathrm{T} (1-V(R+\Delta R)^2)} \, .
\end{equation}
It may be shown by a few lines of algebra that this condition may be
fulfilled only if
\begin{equation}\label{Vc}
V^2 > V_\mathrm{c}^2 = \frac{1}{1+\frac{3}{4s}} \, .
\end{equation}
The analogous derivation for $\Delta R<0$ leads to the same result.
\ignore{
For a power law index for the pressure of $s=2$ (1,3), the minimum
circular velocity for stable dynamical equilibrium is therefore $\sqrt{8/11}$
($\sqrt{4/7}$, $\sqrt{4/5}$) of the Kepler velocity. The corresponding minimum
column density is 
\begin{equation}\label{eq:Nmin}
N_\mathrm{min}= (3+4s) \frac{l}{2\sigma_\mathrm{T}} \,  ,
\end{equation}
which evaluates to 
$N_\mathrm{min}= 8$~($5,11$)~$\times 10^{24} l$~cm$^{-2}$ if $s=2$
(1,3). }
If the rotational velocity is below this value, positive
perturbations to $R$ lead to ejection. 
Negative ones lead to highly eccentric orbits.
We show this by considering the effective potential $V_\mathrm{eff}$.
Inserting eqs.~(\ref{dN}) and~(\ref{dV}) into eq.~(\ref{eq:focl}), and
defining $R_0$ to be a fiducial \rv{distance to the black hole} where eq.~\ref{Neq} holds, with
$N(R_0)=N_0$, $V(R_0)=V_0$ and $x=R/R_0$, we
derive:
\begin{eqnarray}\label{eq:Veff}
V_\mathrm{eff}(x) &=&
 \frac{G M_\mathrm{BH} m}{R_0} \left[ -(1-V_0^2){\cal R}(x)  + \frac{V_0^2}{2x^2} - \frac{1}{x} \right] \\
{\cal R}(x) &=&  \nonumber
 \left\{\begin{array}{ll}\log (x) & s=3/2 \\  \frac{3}{2s-3}x^{2s/3-1}&s\ne 3/2\end{array}\right.
\end{eqnarray}
\begin{figure}
\centering
\includegraphics[width=.47\textwidth]{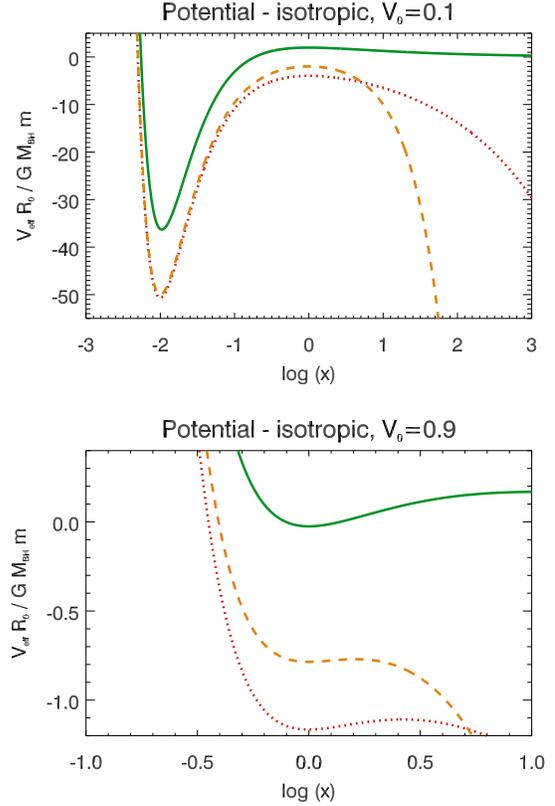}
\caption{\small  Normalised effective potential against normalised
  \rv{distance between cloud and black hole}  for a rotation
  \rv{velocity} of \rv{$V_0=0.1$ (top) and $V_0=0.9$} (bottom)
  \rv{normalised to}
  the Kepler value at $R=R_0$ ($x=1$) for the case of isotropic
  illumination. In each case, the solid green (dotted red, dashed
  yellow) line is for $s=1$ (2,3).
}
\label{fig:Veff}
\end{figure}
The effective potential is displayed in Fig.~\ref{fig:Veff}. In
agreement with the preceding discussion, there is an extremum at
$x=1$, whose character depends on $V_0$: For $V_0>V_\mathrm{c}$
(compare eq.~\ref{Vc}), it is a minimum. Stable bound orbits with small
radial motions may be found in this case.  For $V_0<V_\mathrm{c}$,
the extremum at $x=1$ is a maximum. In this case, their exists a
minimum further in. The equations can be solved easily analytically
for $s=0$, $3/2$ and 3. For $s=3/2$, the second extremum is at:
\begin{equation}\label{xe2}
x_\mathrm{e2}=\frac{V_0^2}{1-V_0^2} \, .
\end{equation} 
This extremum is a
minimum for $V_0<V_\mathrm{c}$. We have verified numerically that for all $1<s<3$, $x_\mathrm{e2}$ is
very close to the value of the $s=3/2$-case, for $V_0<V_\mathrm{c}$. Again, stable bound solutions may be
found. If a cloud is very deep in that potential well, the orbits are
close to circular, and dominated by rotation. The average column
density and rotation velocity for
such an orbit are much higher than the values at $R_0$. Such orbits
are therefore effectively on the stable, solid black part of the equilibrium
curve in Fig.~\ref{fig:dec_N}.
Orbits with total energy close to the potential energy at the maximum
will follow highly eccentric orbits, with predominantly radial
kinematics. For a significant fraction of their orbital period, they
have indeed low column densities and rotational velocities,
corresponding to the region above the red dashed part of the line in Fig.~\ref{fig:dec_N}.

\subsection{Anisotropic light source}
The luminosity in Eddington units for a geometrically thin accretion
disk is a function of the
polar angle and is given by $2 l |\cos \theta |$, where $l$ is the total luminosity of
the source in Eddington units. The force is still central and therefore, 
angular momentum is conserved.
The orbits are planar. Any particular orbital plane may
be characterised by a polar angle $\theta_\mathrm{o}$. We define the
direction of the maximum elevation above the equatorial plane to have
an azimuth of $\phi=0$. A cloud will then have maximum radiation
pressure support at $\phi=0$ and $\phi=\pi$, and no radiation pressure
support at $\phi=\pi/2$ and $\phi=3\pi/2$, when passing the equatorial
plane.  Geometrical considerations result in: $\cos \theta = \cos \phi
\, \cos \theta_\mathrm{o}$. Using this, and eqs~(\ref{eq:focl}),~(\ref{dN}) and~(\ref{dV}),
results in the following total force equation:
\begin{equation}\label{eq:focl2}
F= \frac{GM_\mathrm{BH} m}{R^2} 
\left( \frac{3 l |\cos \phi
\, \cos \theta_\mathrm{o}|}{\sigma_\mathrm{T} N_0 x^{-2s/3}} + V_0^2 x^{-1} -1\right) \, ,
\end{equation}
where $x=R/R_0$, $R(\phi=\pi/2)=R_0$, $N(\phi=\pi/2)=N_0$. The
azimuthal velocity at $\phi=\pi/2$ in Kepler units is denoted by
$V_0$.
Because of the $\phi$-dependence, the force is no longer
conservative. We therefore expect the total energy and hence also the
major axis of the orbits to grow or diminish, depending
on whether the cloud moves predominantly with or against the radiation
flux.  

\begin{figure}
\centering
\includegraphics[width=.47\textwidth]{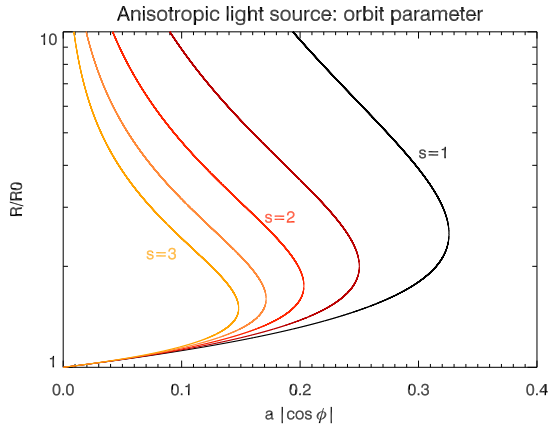}
\caption{\small  Force-free lines for the case of the anisotropic
  light source: \rv{normalised distance between cloud and black hole} versus $a |\cos \phi|$ for different
  values of the pressure distribution index $s$. The force is outward 
to the right of each curve, and inward on the left side. Around the lower
branch of each line, the force is always restoring. Stable orbits are therefore expected 
in this region of the parameter space. If the cloud has low enough column density,
a \rv{too} large part of the orbit will be right of the force-free line, where the force is always 
outward\rv{, leading to ejection of the cloud.}
}
\label{fig:R-a}
\end{figure}

First, we consider the force free locations,  which are useful to
understand the general structure of the allowed orbits. 
In the standard treatment of the 
Kepler problem, bound orbits require the existence of
a local minimum of the effective potential. The elliptic orbits are then oscillating
around this minimum. The forces in this case vanish, of course, at the minimum. Away 
from the minimum, there is an effective restoring force. 
If the total energy is too large, the object can reach a region where the effective force 
is away from the equilibrium point, and may escape.
A similar analysis may be done for the present problem:
Making the definition:
\begin{equation}\label{adef}
a=\frac{3 l \cos \theta_\mathrm{o}}{\sigma_\mathrm{T}N_0} \, ,
\end{equation}
we find the condition for force-free lines:
\begin{equation}\label{eq:peanut}
a |\cos \phi| = x^{2s/3} (1- V_0 x^{-1}) \, .
\end{equation}
A force-free line in the orbit plane is defined by specifying the
parameters $a$ and $s$. Since there is no radiation force in the
equatorial plane, force-free lines require $V_0=1$.
We show $x$
as a function of $a\, |\cos \phi|$ for different values of $s$ in Fig.~\ref{fig:R-a}.
Each curve is characterised by a rightmost value, 
$a\, |\cos \phi| =a_\mathrm{max}$, where the upper branches with negative
slopes and the lower branches with positive slopes meet.
The force is restoring in the vicinity of the lower branches with positive slope.
Therefore, if $a\, \cos \phi <a_\mathrm{max} \forall \phi$ , we expect
bound orbits to exist. We do not necessarily expect bound orbits if
$a\, \cos \phi$ exceeds this value for some fraction of the orbit. 
From eq.~(\ref{eq:peanut}), we find:
\begin{equation}
a_\mathrm{max} = \frac{3}{2s}
\left(\frac{1}{1+\frac{3}{2s}}\right)^{\frac{2s}{3}+1} \, .
\end{equation}
For $s=2$ (1,3), $a_\mathrm{max}$ evaluates to 0.20 (0.33, 0.15). 
Consequently, there is a $\theta_\mathrm{o}$ and $l$-dependent critical column
density, above which bound  orbits should be found: 
\begin{equation}\label{eq:Nmin2}
N_0 \gtrsim 7 \times 10^{23} \mathrm{cm}^{-2} \frac{l}{0.1} \frac{\cos \theta_\mathrm{o}}{0.5}
\frac{0.33}{a_\mathrm{max}}\, .
\end{equation}

\begin{figure*}
\centering
\includegraphics[width=.32\textwidth]{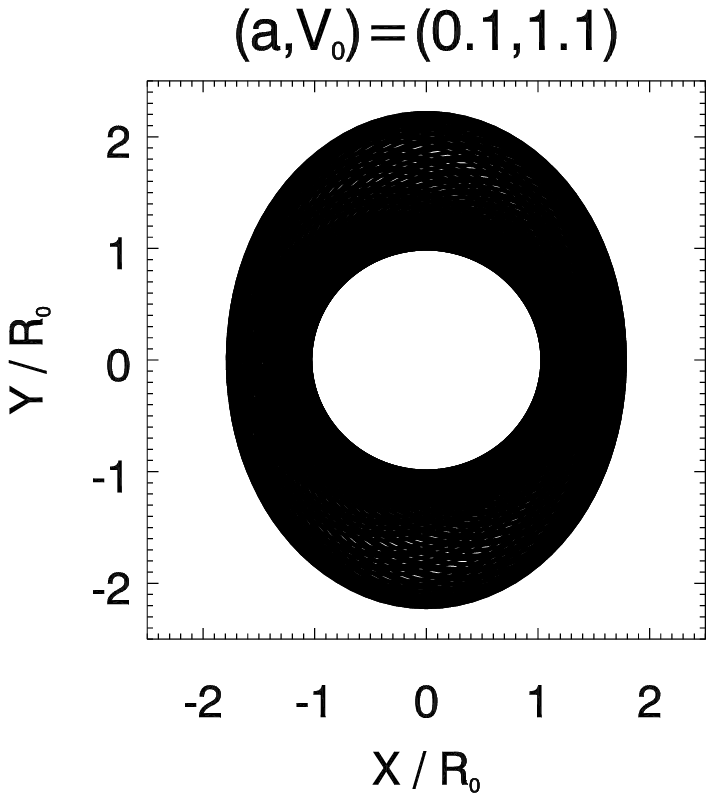}
\includegraphics[width=.32\textwidth]{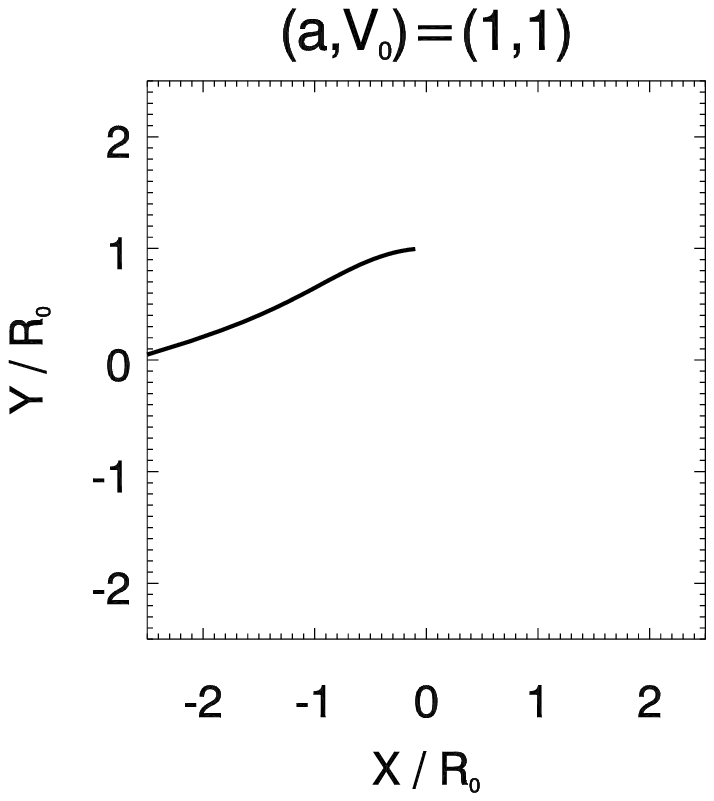}
\includegraphics[width=.32\textwidth]{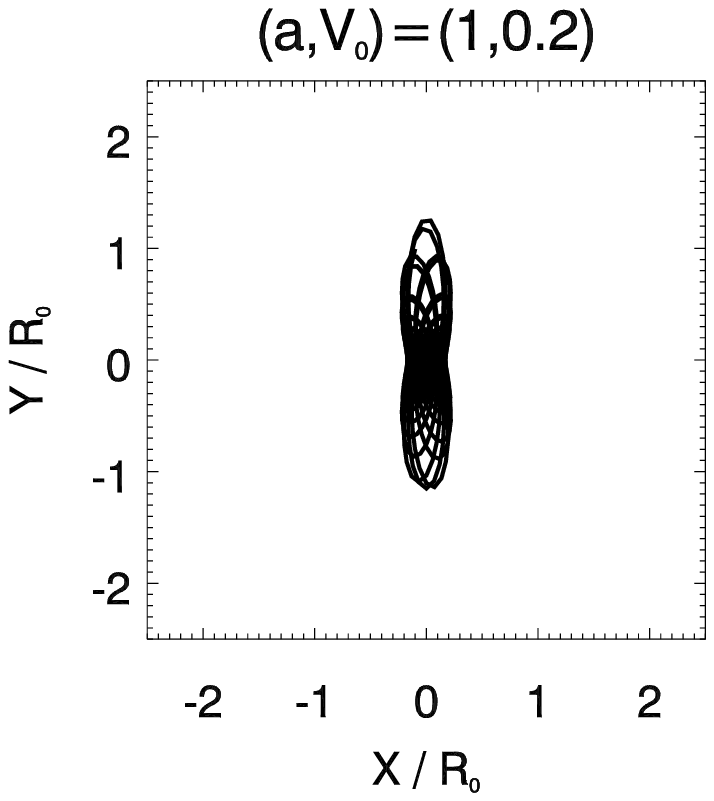}
\mbox{}\hspace{2.5cm}\mbox{}\hspace{\fill}$(a,V_0)=(0.1,1.1)$\mbox{}\hspace{7cm}\mbox{}$(a,V_0)=(1,0.2)$\hspace{\fill}\mbox{}\\
\includegraphics[width=.27\textwidth]{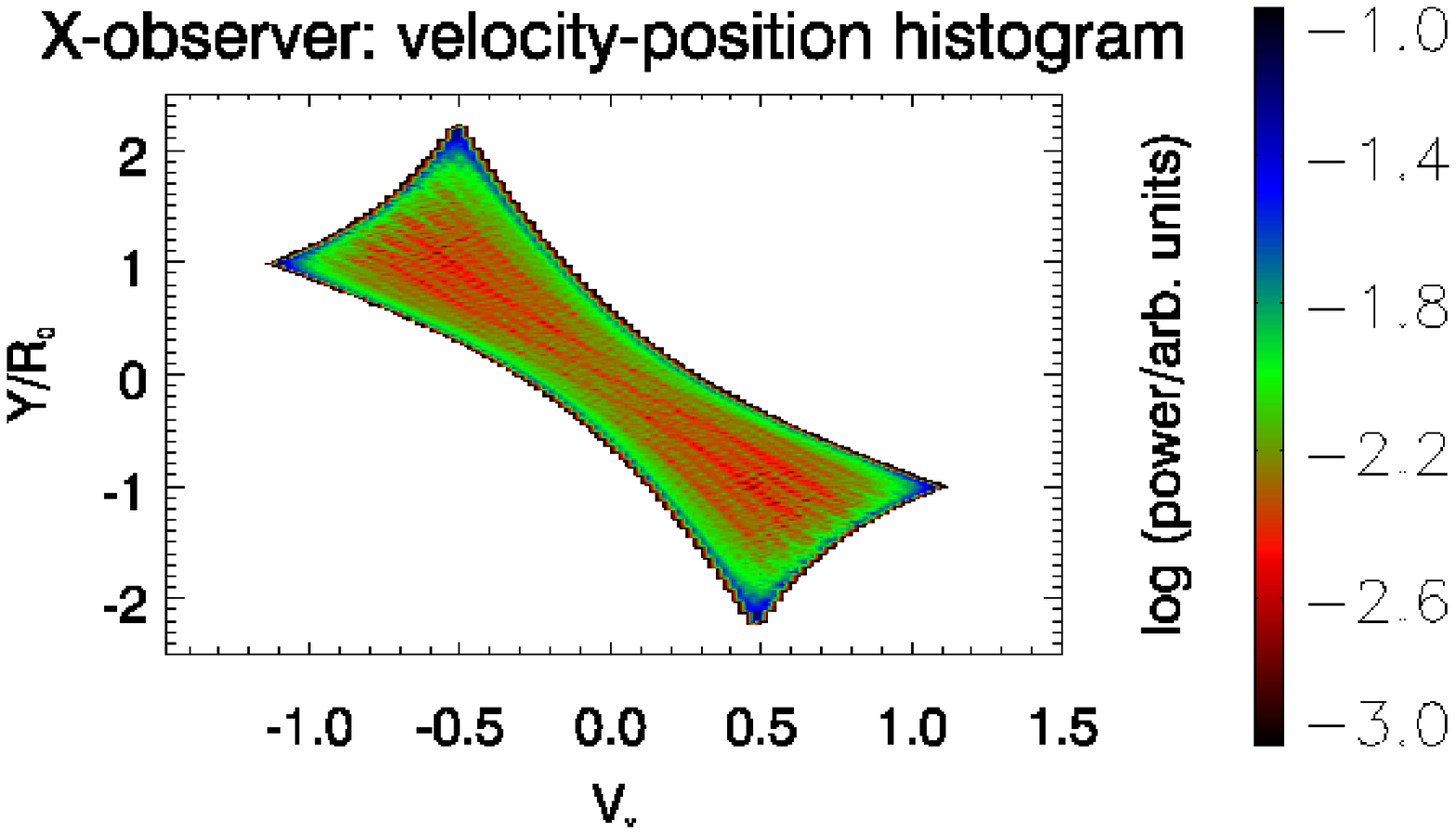}
\includegraphics[width=.22\textwidth]{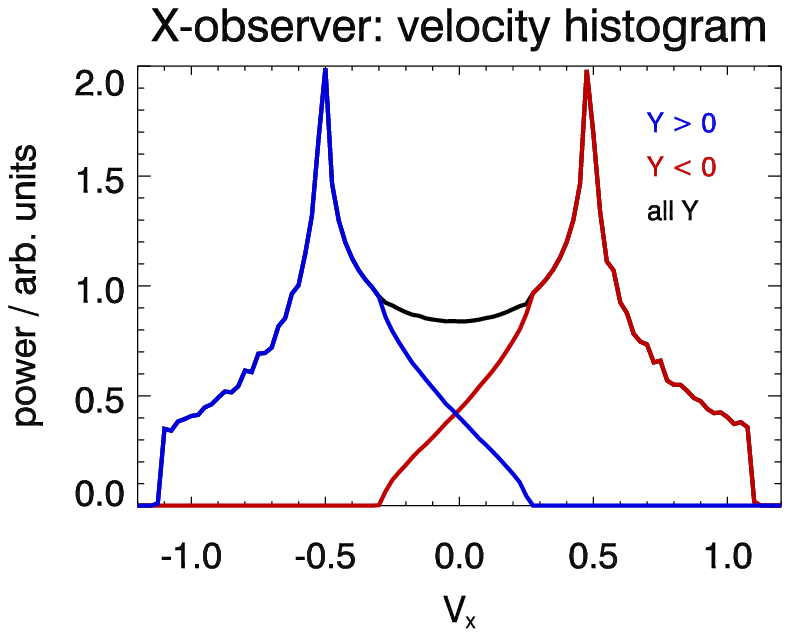}
\includegraphics[width=.27\textwidth]{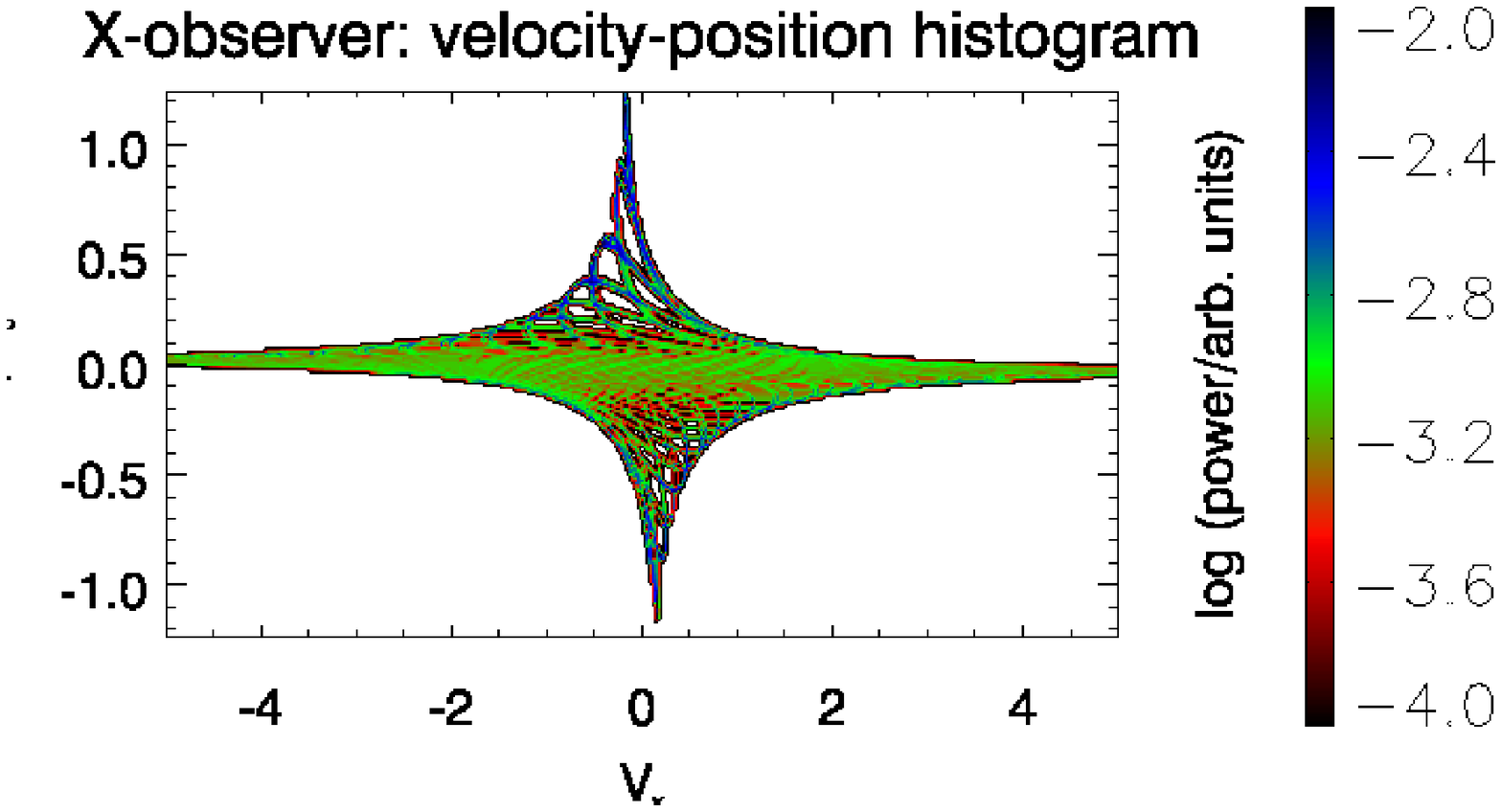}
\includegraphics[width=.22\textwidth]{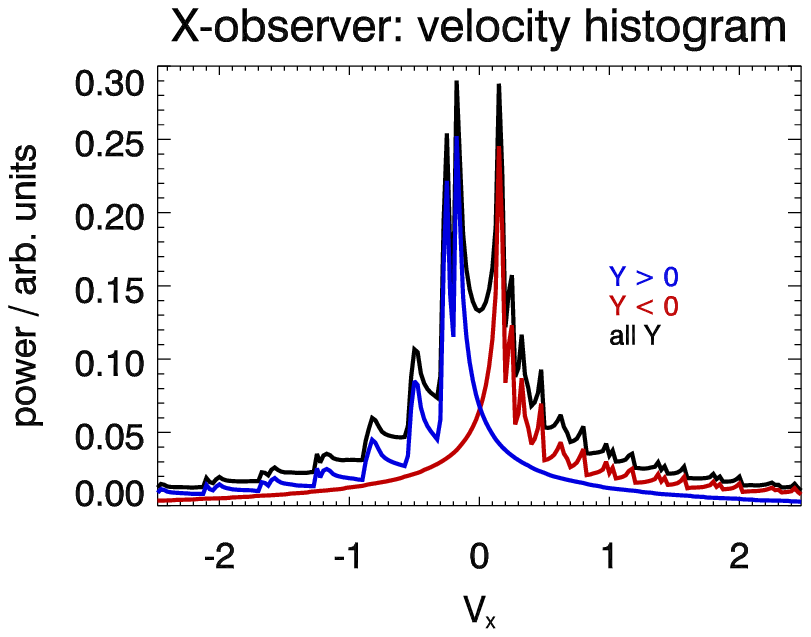}
\includegraphics[width=.27\textwidth]{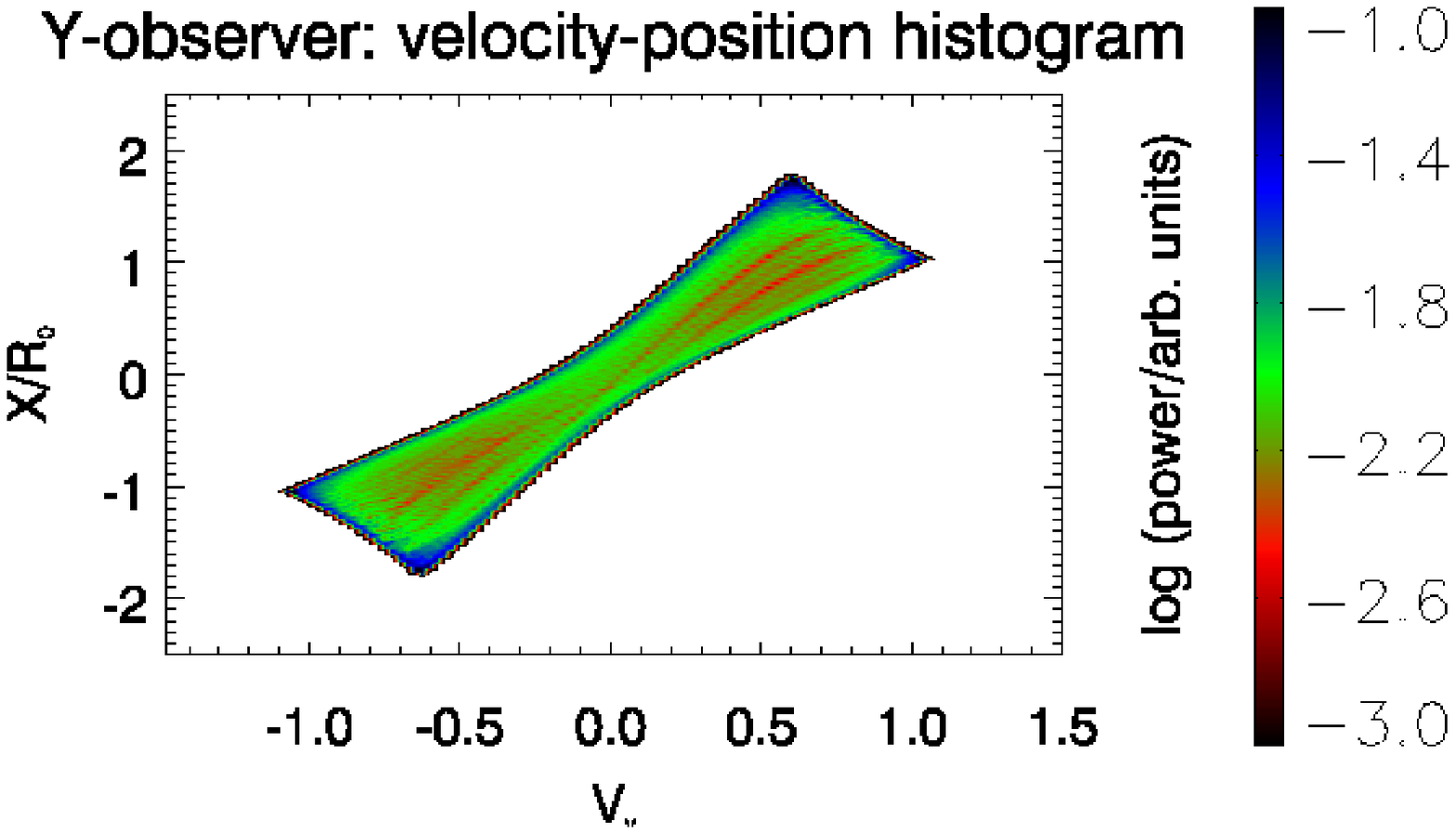}
\includegraphics[width=.22\textwidth]{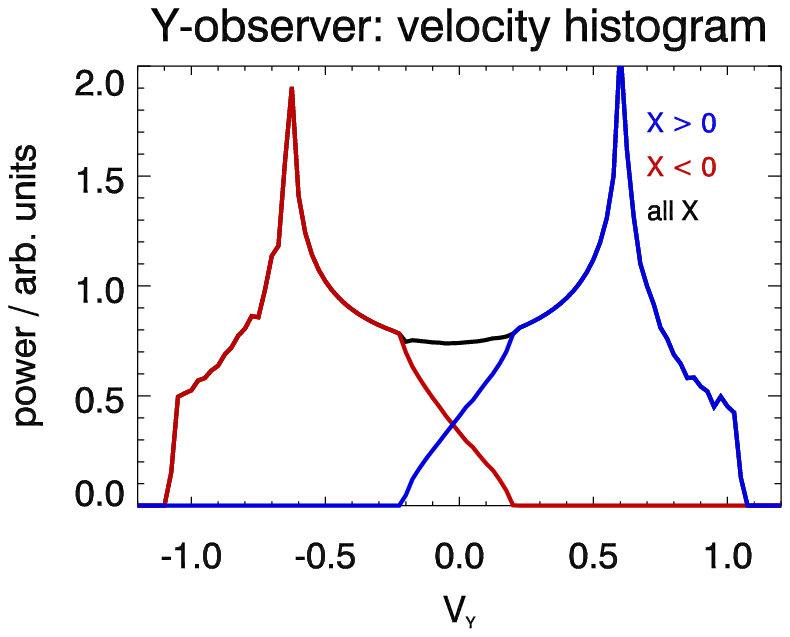}
\includegraphics[width=.27\textwidth]{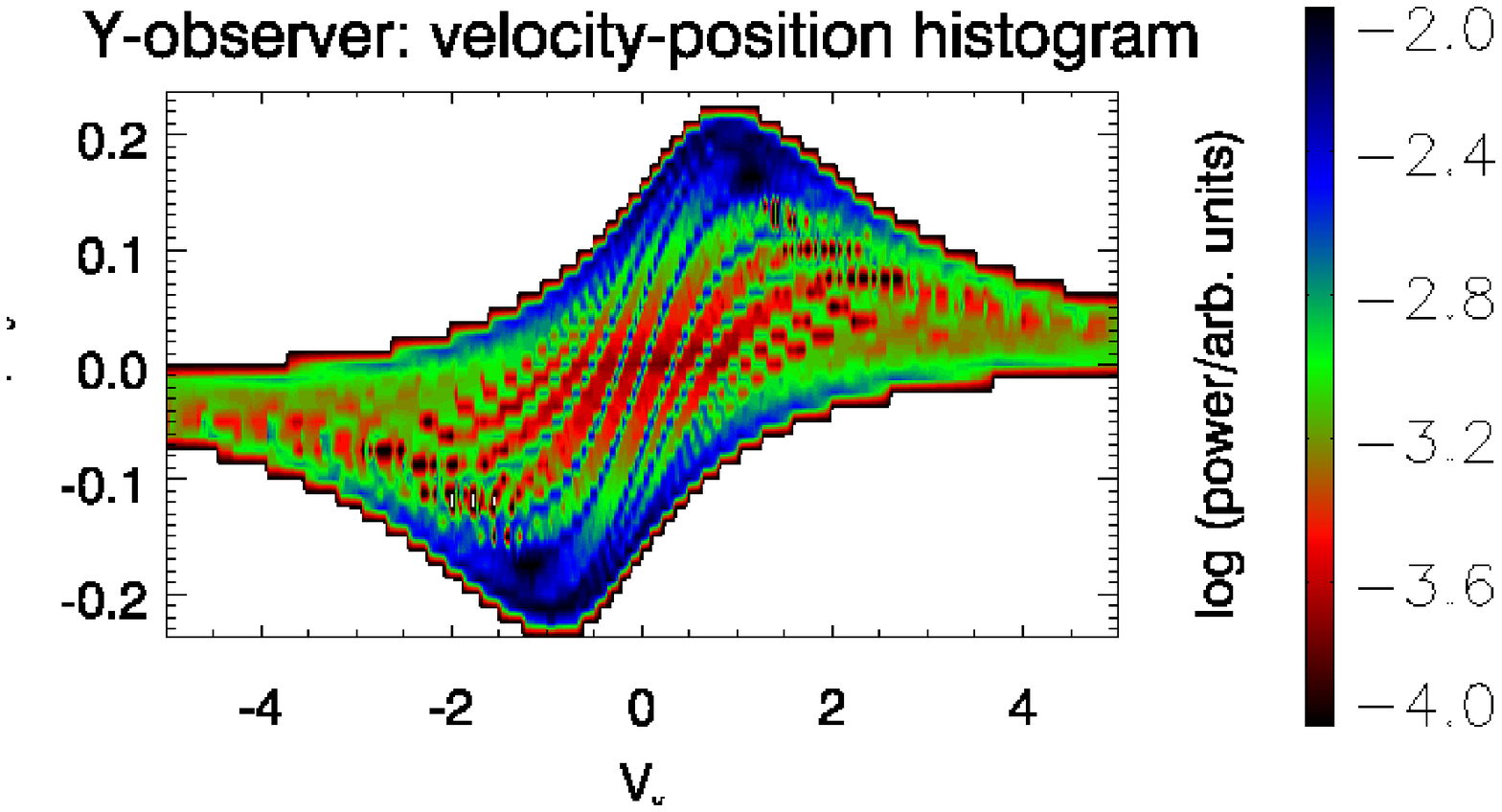}
\includegraphics[width=.22\textwidth]{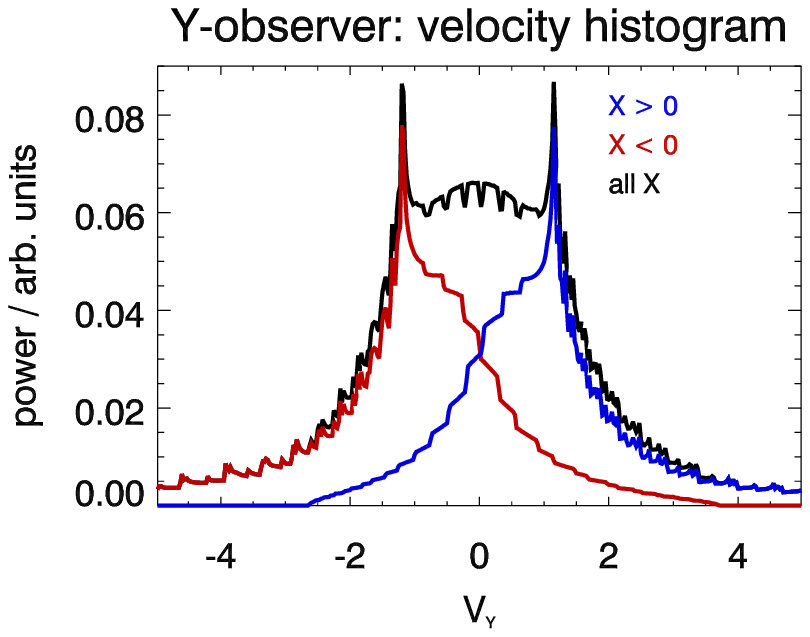}
\caption{\small  Position and kinematic data for three example orbits
  for the anisotropic case. Radiation pressure support is maximal on the X-axis, and
  zero on the Y-axis. We have integrated for 100 orbits, for the bound
  cases. The simulations are converged with regard to the
  timestep. Top row: orbit shapes in orbit plane. 
  \rv{The orbits} are similar to
  precessing Kepler-ellipses. Because of the non-conservative nature of the
  potential, they grow and shrink periodically, with a period of
  $\pi$. This leads to Rosetta shapes, which are compressed in the
  direction of maximum radiation support. The set of four plots on the
  lower left shows kinematic information for the \rv{$(a,V_0)=(0.1,1.1)$}
  simulation. Here we assume that a cloud emits radiation on the
  illuminated side facing the centre. We calculate the fraction of
  that surface seen by respective observers. The emission is further
  proportional to the current cloud surface and the time interval.
  The top row relates to an observer in $-X$-direction \rv{(X-observer)},
  the bottom one to an observer in $-Y$-direction \rv{(Y-observer)}. The left column shows
  position-velocity diagrams, the right one the emission at a given
  velocity \rv{(black) and separately for negative (red) and positive
    (blue) values of the respective complementary coordinate}. \rv{The
  emission of the two peaks at about half the Kepler velocity is
  easily discerned.} The
  lower right set of four plots shows the same for the run with
  \rv{$(a,V_0)=(1,0.2)$. The emission occurs predominantly at
    smaller velocities. In velocity space, emission from the sides with positive
    coordinates still has little overlap with the emission from the
    other sides.} Colour scales are cut at reasonable values, as indicated
  at the individual colour bars. \rv{Velocity is given in units
  of the local Kepler velocity at the initial cloud position $R_0$.}
}
\label{fig:orbs}
\end{figure*}

Similarly as in the isotropic case, we also find bound, but
highly eccentric orbits
for low column densities and rotation velocities:
We show this by numerical integration of some example orbits
(Fig~\ref{fig:orbs}). 
\rv{Our reference in the following is a
  Cartesian coordinate system with coordinates $X$ and $Y$ defined in the
  orbital plane. The $Y$-axis is taken to be the intersection between
  the orbital and the equatorial plane.} 
We use $s=2$ and start the clouds at $\phi=\pi/2$, i.e. where
the orbital plane meets the equatorial plane (positive $Y$-axis in 
Fig~\ref{fig:orbs}), with certain values for
$a$ and $V_0$. For $(a,V_0)=(0.1,1.1)$,
we expect a stable bound orbit despite the super-Keplerian initial
velocity, since $a$ is in the stable regime. This is indeed the case
(Fig.~\ref{fig:orbs}, top left). For $(a,V_0)=(1,1)$, we expect ejection,
since $a$ is in the unstable regime and the comparatively high
velocity corresponds to a positive radial perturbation. Again, this is
what we find (Fig.~\ref{fig:orbs}, top middle). For $(a,V_0)=(1,0.2)$, we
expect highly eccentric orbits. This is confirmed by the numerical
integration (Fig.~\ref{fig:orbs}, top right). Here, the non-conservative
nature of the potential is most apparent: The cloud gains energy, when
moving outwards. But since it advances in azimuth, it does not get
back the same amount on the way inwards. However, on average over
many orbits, the contributions cancel each other. 

In order to obtain kinematic information, we also recorded the 
emission measure, $e$, of the cloud into a certain
direction during the integration, together with the current position
and velocity. At each time $t$, it is
calculated as:
\begin{displaymath}
e(t) = A(\phi) \, \mathrm{d}t \, R^{2s/3-2} \, ,
\end{displaymath} 
where $A(\phi)$ is the fraction of the illuminated cloud surface seen
by the observer. \rv{We take only the side facing the AGN as line emitting
region, as appropriate for an optically thick line.
The current time step interval is denoted by $\mathrm{d}t$,} and
the radial dependence is due to the change of the cloud size and
radiation flux with distance to the centre.
\rv{We place two observers at large negative $X$ and $Y$ values,
  respectively. At each timestep, we project the clouds velocity along
  the respective lines-of-sight. $e(t)\,\mathrm{d}t$ is then added to
  the corresponding bins of line-of-sight velocity and transverse
  position in order to create a two-dimensional emission-weighted histogram. }

\rv{For the bound orbits, these velocity resolved emission-weighted
  histograms are  shown in
Fig~\ref{fig:orbs}.} A priori, one might expect a
\rv{broader signal} for the low $a$ (high column density)
case \rv{with well separated emission peaks near the positive and negative
Kepler velocity}, and a \rv{narrower signal} for the high $a$
(low column density) case. This is indeed what we find. For
$(a,V_0)=(0.1,1.1)$, the peak of the emission is at \rv{$\pm\approx
  50$~per cent of the
Keplerian velocity at $R_0$.}
For $(a,V_0)=(1,0.2)$, the orbits  get very anisotropic in real space as well
as in velocity space. An observer in the $-X$-direction would see two
peaks at around \rv{$\pm 0.15$} times the Kepler velocity at $R_0$, close to
the initially imposed one. Here, the outermost locations of the orbit
dominate the emission due to the longer time spent there.
An observer in the $-Y$-direction would see the two peaks at around
$1.2$ times the Kepler value, because from this point of view, the
orbits are much \rv{narrower. If}
there was an ensemble of such clouds with the angular momentum vector
randomly rotated around the symmetry axis of the system, but otherwise
identical, the signal at low velocity would dominate, as the
emissivity at the slower peak is about three times higher. For an
observer who would see the orbital plane at some inclination, the
apparent velocities would be somewhat below these values.
\rv{Since spectropolarimetric observations are able to resolve the BLR
in many objects, we have also separated the emission that comes from
the positive part of the transverse axis from the one of the
opposite side. As one might have expected, the emission from the two
sides is well separated in velocity space, in both cases. Remarkably,
for the emission from a given side, the $(a,V_0)=(1,0.2)$ cloud shows
about ten percent of the peak emission of the opposite side at the
location of the peak of the opposite side. In contrast, for the cloud
with $(a,V_0)=(0.1,1.1)$, the emission of a given side drops to zero
at the peak of the other side.}

\section{Discussion}\label{disc}
We have shown that pressure confined clouds at any sub-Keplerian rotational velocity
may exist in stable
dynamical equilibrium in the BLR. 
Given the significant evidence for a
gravitationally bound, flattened, but still of considerable thickness,  
and disk-like geometry for the BLR
(compare sect.~\ref{intro}), it is reasonable to require a stable
dynamical equilibrium for the line-emitting clouds. 

An essential
ingredient for the model is the behaviour of the pressure of the
inter-cloud material with radius:
A reasonable assumption for the inter-cloud component is an 
Advection Dominated Accretion Flow (ADAF, e.g. \citeauthor{NY94}
\citeyear{NY94}; \citeauthor*{YMN08} \citeyear{YMN08}).
A nearly hydrostatic solution is included in the ADAF models for the limit of
low accretion rates. For this type of solutions, the power law index
for the pressure $s$ is between two and three. Outflow solutions have
also been considered in the literature. \citet{KK94} find $1<s<1.5$.
Therefore, $s$ should be between one and three
(similar results are obtained by \citeauthor*{RNF89} \citeyear{RNF89}).

We have considered an isotropic and an anisotropic light source as 
appropriate for an accretion disk. The former is usually implied in
the literature. For the isotropic case, the force is central
(i.e. conserves angular momentum) and conservative. We find an
equilibrium relation between column density, luminosity of the AGN and
rotational velocity. We show by direct stability analysis that only \rv{for}
the part of that relation with high rotational velocities \rv{a cloud
  would encounter a restoring force for small radial perturbations.} However, by analysis
of the effective potential, we show that stable orbits may be found
also for low rotational velocities and column densities slightly above
the equilibrium curve given by eq.~(\ref{Neq}).  However, the character of the\rv{se}
orbits change\rv{s}: while high column densities (corresponding to close to
Keplerian rotation) allow for even circular orbits, low column density clouds require highly
eccentric ones.

For anisotropic illumination, the force is still central, and
therefore angular momentum conserving, but it is no longer
conservative. Therefore, as the orbits precess, there is now the
additional feature that they gain and loose
total energy, which is exchanged with the radiation field. This is
evident from the periodic change of the major axis of the orbits. We show
that the time averaged signal of such a cloud would also be very
anisotropic. If one would consider many such clouds with
randomly rotated angular momentum vector around the axis of symmetry,
the stronger emission from further out would dominate. In our example,
this produces a peak in emission at a small fraction of the Keplerian
velocity with a similar FWHM. More important, the emission
contributions from the two sides of the accretion disk are
kinematically clearly distinct. This would therefore probably still be compatible with the
spectropolarimetric results. More detailed comparisons using realistic
cloud samples have to be done in order to decide if the dispersion
would be low enough to fit with all the available data. This is
however beyond the scope of this article.

Cloud column density and rotation velocity in the anisotropic case are still related by
eq.~(\ref{Neq}) (also Fig.~\ref{fig:dec_N}), when orbit averaged
values are used. Therefore, it is well possible to observe arbitrarily
low rotation velocities for cloud column densities of order $10^{23}
(l/0.1)$~cm$^{-2}$, confirming the results of  \citet{Marcea08}.

Interestingly, it has turned out that for the anisotropic case, the
orbits are \rv{less extended} in the direction of the strongest radiative
force. This might first appear counter-intuitive, but is readily
explained if one considers that the radiative force adds energy to the
orbit, as long as the motion is outward, but brakes down the cloud,
when it moves inward. If the major orbit axis coincides with the
direction of maximum radiation force, the contributions nearly cancel,
whereas one may get positive changes when the major axis has
advanced past that direction. Energy losses are expected, when the
major axis has not yet reached the maximum force direction.

\rv{If the column densities are not too high, the radiation field
  favours a disk configuration:
  Consider an initially isotropic ensemble of clouds with relatively
  low column
  densities, comparable to the one given in eq.~(\ref{eq:Nmin2}), and 
  a broad distribution of angular momenta. 
  Due to angular momentum conservation, each cloud orbits
  the black hole in a particular orbital plane on an elliptical
  orbit. Once an anisotropic AGN is switched on, we may assign an $a$
  value to each cloud. Because $a \propto \cos(\theta_\mathrm{o})$, clouds at low polar
  angle (closer to the symmetry axis of the emission of the central
  accretion disk) have greater $a$ values. If this $a$ value is too
  large, the high angular momentum clouds will be ejected. For the
  remaining clouds at low polar angle, the major axis of their orbits will shrink
  when it points towards higher latitudes on their course of
  precession. In Fig~\ref{fig:orbs} we have demonstrated this
  effect. The result will be a disk-like BLR.
For column densities much higher than the one given in
eq.~(\ref{eq:Nmin2}), 
the orbits are less
affected by radiation pressure. For such clouds, t}he BLR would therefore not be
constrained to have a disk or other geometry by these dynamic
considerations. \rv{It may or may not be in a disk configuration for
  other reasons. The critical column density is of the order
  $10^{24}$~cm$^{-2}$ for common luminosities of ten per cent of the
  Eddington value, which is rather large
  compared to observational constraints (compare references in
  sect.~\ref{intro}). One might therefore generally expect this effect
to be significant in many BLRs.}
 
The detailed mixture of orbits is of course very hard to
predict. Since the cloud mass is unimportant for the acceleration of
the cloud, we expect clouds of a wide range of masses, with the column
density adjusting according to the \rv{cloud's} position in phase
space.
\rv{Some BLR clouds might have been born in situ
  \citep[e.g.][]{PD85}.} If the clouds would have come from further out, an
interaction would be required to reach the bound orbits. This might
favour eccentric orbits. The mixture of orbits will determine the
observed velocity structure.

Pressure confined spherical clouds as used in this analysis, suffer from
shearing by differential radiative forces due to the varying column
density from the cloud's rim to it's centre \citep{Mat86}. For our
parameters one would expect complete disruption after about
 a hundredth of an orbital period. This issue is common to this class
 of cloud model \citep[compare e.g.][]{RNF89}, and has not been solved
 so far. Possible ideas to stabilise the clouds include magnetic
 fields and a more favourable geometry \citep[e.g.][]{Net08}.

\section{Conclusions}\label{conc}
We have shown that pressure confined clouds may rotate stably on bound
orbits near the dynamical equilibrium between radiation, centrifugal
and gravitational forces at all sub-Keplerian rotational velocities. 
This is true for isotropic illumination, as
well as for the case where the radiation flux is correlated
with the polar angle. \rv{While angular momentum is conserved in both
cases, energy is not conserved for anisotropic illumination. This
leads to Rosetta orbits that extend less in the direction of maximum
radiation force. An intrinsically isotropic low column density
cloud system would therefore become less extended in the polar
directions, when an anisotropic AGN would be switched on, and
consequently appear disk-like.} We show that it is possible
to find clouds \rv{of low and high rotational velocity with well separated peaks in the
spatially resolved emission spectra} as a function of velocity, 
as required by spectropolarimetric
BLR data. These findings confirm the
idea that significant corrections of black hole masses due to radiative
forces are possible in certain objects, as proposed by
\citep{Marcea08}. 

\section*{Acknowledgements}
We thank Jim Pringle for very helpful discussions and the anonymous
referee for many useful comments that significantly improved
especially the clarity of the manuscript.

\bsp
\bibliographystyle{mn2e}
\bibliography{/Users/mkrause/texinput/references}

\label{lastpage}


\end{document}